\documentclass[runningheads]{llncs}
\usepackage[utf8]{inputenc}
\usepackage[english]{babel}
\usepackage[T1]{fontenc}
\usepackage{comment}
\usepackage{graphicx}
\usepackage{blindtext}

\begin{document}

\title{TecCoBot: Technology-aided support for self-regulated learning
\thanks{
This work was supported by the German Federal Ministry of Education and Research for the tech4comp project under grant No 16DHB2102
}}
\subtitle{Automatic feedback on writing tasks with chatbots}

\titlerunning{Supporting students in reading texts}
\author{Norbert Pengel\inst{1}\orcidID{0000-0002-3263-6877} \and
Anne Martin\inst{1}\orcidID{0000-0001-8237-6770} \and
Roy Meissner\inst{1}\orcidID{0000-0003-4193-8209}
Tamar Arndt\inst{1}\orcidID{0000-0001-8170-3346} \and
Alexander Tobias Neumann\inst{2}\orcidID{0000-0002-9210-5226} \and
Peter de Lange\inst{2}\orcidID{0000-0002-3494-7513} \and
Heinz-Werner Wollersheim\inst{1}\orcidID{0000-0002-4690-5839}}

\authorrunning{Norbert Pengel et al.}

\institute{Leipzig University, Faculty of Education, Leipzig, Germany,\\
\email {[norbert.pengel,anne.martin,roy.meissner,tamar.arndt, wollersheim]@uni-leipzig.de}
\and
RWTH Aachen University, Aachen, Germany\\
\email{[neumann,lange]@dbis.rwth-aachen.de}
}

\maketitle

\begin{abstract}
In addition to formal learning at universities, like in lecture halls and seminar rooms, students are regularly confronted with self-study activities. Instead of being left to their own devices, students might benefit from a proper design of such activities, including pedagogical interventions. Such designs can increase the degree of activity and the contribution of self-study activities to the achievement of learning outcomes.

Especially in times of a global pandemic, self-study activities are increasingly executed at home, where students already use technology-enhanced materials, processes, and digital platforms. Thus we pick up these building blocks and introduce TecCoBot within this paper. TecCoBot is not only a chatbot, supporting students in reading texts by offering writing assignments and providing automated feedback on these, but also implements a design for self-study activities, typically only offered to a few students as face-to-face mentoring.

\end{abstract}

\section{Introduction}
\label{sec_intro}




Self-study activities are an integral part of today's workload in higher education \cite{EUECTS2015}. However, studies on the workload of students show that self-study activities plays a subordinate role and that too little time is spent on independent learning \cite{Metzger2010}.
One approach that is widely recognised as effective in positively influencing students is mentoring \cite{Dolan2015}. However, individual face-to-face mentoring requires a lot of time and effort. Especially in higher education, where a large number of students is often supervised by only a small number of lecturers, it is impossible to provide holistic mentoring to all students.

As part of the German research project tech4comp, which searches for design concepts to scale mentoring activities by utilising technology, we developed a chatbot that is able to provide informative feedback for a large number of students fast, as part of self-study activities.
This is even more important, as students currently cannot be at a university due the corona pandemic 2020 and thus cannot make use of face-to-face mentoring.




Various models describe the knowledge and research process in (educational) design research. 
Based on a review of existing models of design research, a generic model was developed \cite {McKenney2019}, which contains three core phases in a flexible and iterative structure: 1) analysis/exploration, 2) design/construction, and 3) evaluation/reflection, which we currently use in our work.

In this paper we address the results of the first two phases, while our focus is on a prototype for a technology-aided support system that provides students with automated feedback in form of knowledge graphs on submitted writing assignments via a chatbot. 

\section{Requirements and Related Work}
\label{sec_RW}
 
 \subsubsection{Evaluating student needs} 

In accordance with the multi-method approach, we use both qualitative and quantitative methods.
We used the “Teaching Analysis Poll (TAP)” as a qualitative method for exploring students needs in terms of learning. Actually it is an instrument for course evaluation due to open-ended questions. 
This method has a low degree of structuring and thus the greatest possible exploratory character of result generation. 
In 2019 296 of 628 enrolled students were surveyed in an educational science seminar in teacher education at Leipzig University.
The test persons were asked about the parts of teaching-learning setting and activities of teaching staff of universities that either promoted or hindered learning. 
The evaluation of the results was also carried out with a qualitative method of empirical social research, the qualitative content analysis in direct coding on the data material. A coding guideline prepared for TAP\footnote{guideline:\url{https://https://epub.uni-regensburg.de/35604/}} served as a structuring aid for this.
The structuring analysis was carried out in several rounds in which the data material was compared with the individual categories.
These TAPs took place in all seminar groups in the above mentioned period and offered qualitative intermediate feedback on the courses, which was already condensed to statements that were accepted by a majority in a Think-Pair-Share procedure. 
Altogether, a clear picture of the learning situation as perceived by the students could be derived. 
In addition to the large number of texts to be read in the seminars, lack of focus on the content and its structure, which is difficult to recognise, were mentioned as obstacles to learning.
Following this poll, a quantitative survey with a closed questionnaire is planned.
The aim of this survey is again to determine students' need for support and to test hypotheses derived from the qualitative data (in this respect it is a sequential research design). 
Accompanying this, however, the theoretical research that also takes place in this phase is important for structuring the initial situation of research process \cite{Reinmann2014}.
 

\subsubsection{Mentoring}
It is understood as dyadic relationship between mentor and mentee and can be an effective pedagogical measure to support learners in their learning process~\cite{Ziegler2009}. 
A holistic view of mentoring includes cognitive, motivational and emotional dimensions of this interaction, while ensuring learner autonomy and self-determination. 

If mentoring is to be designed to promote and motivate learning, the findings of learning research are important, especially on self-regulated learning and the related metacognitive regulation. 
The greatest effects of mentoring can be expected when optimal conditions for effective learning processes are in place, including frequent and high-quality feedback, practice and consolidation of what has been learned. 
Currently, we focus on mentoring to support domain-specific learning processes, especially knowledge.

In order to systematize the concept of mentoring, which is presented in the research literature as being not very homogeneous, we refer to its functional criteria \cite{Nora2007}, \cite{Gershenfeld2012}. 
These include (among other things) support for the transfer of scientific expertise, which aims to promote the students' knowledge relevant to the chosen subject area.
The development of the prototype will focus on knowledge of the students.

\subsubsection{Generating feedback}
For mentoring to be accepted by students, the quality of feedback is important \cite{Hoeher2014}. 
Findings from empirical educational research clearly show that feedback can have a positive influence on learning and development processes. 
In addition to personal factors of the sender and recipient (e.g. expertise, self-efficacy, attribution patterns) and situational factors (cultural conditions, binding character), most important is to design the transmission of feedback in a way that promotes development of the student \cite{Kluger1996,Timperley2007,Hattie2013}. 
A distinction can be made between informative and controlling feedback \cite{Deci1985}.
The latter impairs intrinsic motivation by building up pressure and thus, according to the theory of self-determination \cite{Deci1985}, a condition for sustainable learning. 
According to a synthesis of meta-analyses of feedback in schools \cite{Hattie1999}, informative feedback is also considered more effective than praise or punishment. Important for the effectiveness of feedback is that it offers learners clues or encouragement and relates directly to learning goals.

Accordingly, we provide writing assignments and automated feedback for accompanied self-study activities. 
The communication of these tasks and feedback are transmitted with a chatbot and is characterized by an understandable everyday language without academic language in order to increase the possibility of reflection and also to stimulate cognitive, metacognitive and motivational strategies. 
Through this intervention communicated by chatbot, students should understand content and integrate it into their existing knowledge structure. 
By activating metacognitive strategies, it should be possible for them to better monitor their own learning process.

\begin{figure}
  \vspace{-1.5em}
  \centering
  \includegraphics[width=0.9\textwidth]{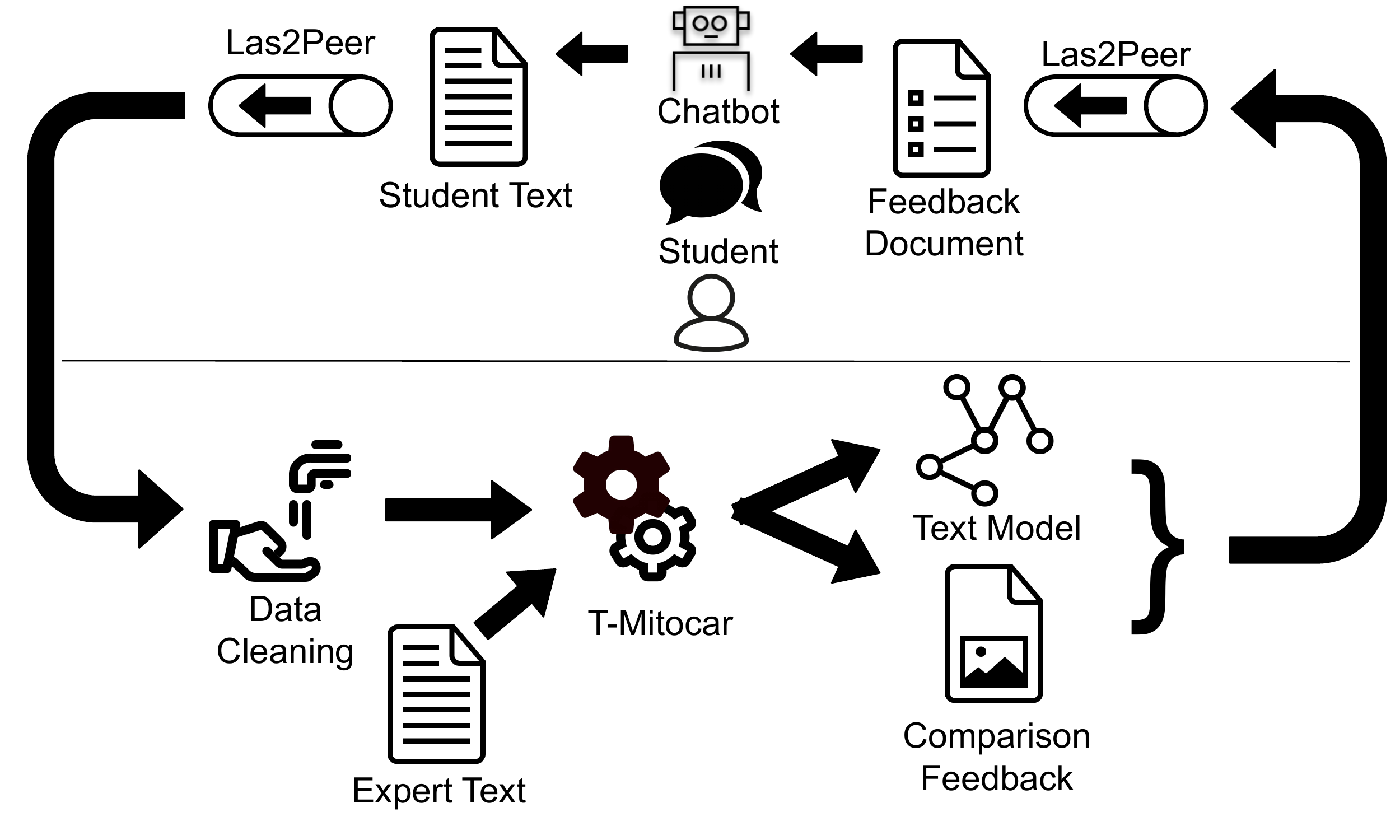}
  \caption{Workflow of the mentoring prototype for text feedback}
  \label{fig2}
  \vspace{-1.5em}
\end{figure}

\section{Approach and Implementation}
\label{sec_AppImpl}

\subsubsection{Feedback with T-MITOCAR}
We are using the software T-MITOCAR to provide individual feedback on writing assignments.
T-MITOCAR constructs re-representations of knowledge from prose text through a computer-linguistic analysis, without the need to incorporate an external knowledge base \cite{Pirnay2015a,Pirnay2015c}. 
The resulting knowledge graphs look similar to mind maps and may be modeled from texts passages, like book chapters, or whole texts, like written by students.
In our case, the knowledge graph of a students text is compared with the graph of reading assignment text (reference text) to generate concrete feedback. 
For instance, this feedback can inform students about which key concepts are part of their text model and also show how these are interlinked. 
It can also inform about which concepts overlap with the model of the reference text, which are different, and about how the concepts are linked in the assignment text. 
It should also be noted that similarity to the assignment text depends on content and on the writing task, so similarity is not always a goal parameter \cite{Aimeuer1998,Helbig2006,Jonassen2006,vandenBroek2010}. 
All this information can aid reflection and text revision and, importantly, in combination with active writing it can foster an in-depth study of the topic in self-regulated learning \cite{Tynjaelae2001}. 

\subsubsection{Feedback provided by Chatbot}
Chatbots are conversational interfaces, allowing humans to interact with software using natural language. 
Therefore chatbots are known to be relatively easy to use and intuitive with a low entry barrier and good accessibility \cite{Holmes2019}, which makes them suitable for university context with users on different levels of digital literacy. 
Classic mentoring typically takes place within conversations between a mentor and mentee. The conversational nature of chatbot interactions thus seems to be an easy choice to digitally emulate mentoring processes. In contrast to most static graphical interfaces a chatbot is inherently adaptive to some degree since it reacts to the users input in the course of conversation, depending on the chatbot's underlying knowledge base. 
To a large extent the efficacy of mentoring lies in its adaptiveness towards the mentee and his or her needs and competences \cite{Vandewaetere2014}. 
For this a chatbot can fulfill an important requirement for technology-aided mentoring although we still need to determine the elements the chatbot needs to adapt to through the course of conversation. 
Other elements known to be key factors for good mentoring are the trust and respect within a mentoring relationship \cite{Alemdag2017}. 
Reportedly chatbots are potentially trusted and felt safe to share information with by their users depending on implementation and design details of the chatbot, like for example its human-likeness, humor or professional appearance \cite{Folstad2018}. 
This as well as other specifics of the conversational design, like ethical factors, transparency, the degree of human-likeness or opportunities for playfulness within the chatbot conversation, still need to be subject to our future consideration.

Incorporating the results of the analysis and exploration phase, first ideas for solutions were generated and implemented as part of a prototype\footnote{\url{https://gitlab.com/Tech4Comp}} This prototype consists of three parts: the chatbot
, a UI component
inserted into the LMS and a service
, creating the knowledge graphs and the comparisons of these graphs.
A first test took place in April 2020. In summer 2020 we will make the chatbot available to students for the first time.
In the following we describe one prototype as result of the design and construction phase.
\subsubsection{Implemented Prototype} 
A first prototype was developed, tackling the requirements and needs described above. 
It is presented to end users, e.g. students, as a special view in a learning management system (LMS), providing a personalized chat interface with a chatbot.
This chatbot was implemented and trained using the open source natural language processing framework RASA\footnote{\url{https://rasa.com}}.

The chatbot's main purpose is to offer writing assignments to students, gather the results and provide feedback. 
Therefore students are able to upload their written texts within the chat. Such individual texts are proxied through a las2peer network (see next paragraph) to a service generating the feedback. 
The first step this service executes is a cleaning process for the issued texts, as the input for further processing steps needs to be in a specific format. 
Subsequently the service involves T-MITOCAR to do two things: 1) request a knowledge graph that visualizes the concepts and concept connections of the students' text, and 2) request a graph comparison of the just created graph with a reference knowledge graph. T-MITOCAR generated this reference graph from another written text, i.e. from a textbook, that contains domain knowledge about the same topic the students' text is about. 
To hand back proper feedback to students, a feedback template is filled with specific comparison results, which is eventually converted to a feedback PDF document. 
Depending on the mentoring scenario, either the students knowledge graph, the reference knowledge graph or the textual comparison of the two knowledge graphs is transferred through the las2peer network back to the chatbot, which finally provides the feedback document via the chat interface to the individual student. This whole process is depicted in figure \ref{fig2}.

The former described frontend is a LMS-Plugin providing a single thread of a RocketChat\footnote{\url{https://rocket.chat/}} instance to end users, implemented as a web-component. 
Besides the actual chat the plugin also handles authentication, authorization, and data protection in accordance to the European GDPR rules. 
Regarding authentication and authorization an OpenId Connect (OIDC) login barrier is used, which makes sure that student credentials stay safe. 
Using OIDC furthermore allows students to use one account to login to the LMS, the chat client, and the las2peer network. The latter one is a decentralized open source environment for transferring and storing user data without inheriting a central authority \cite{klamma2016las2peer}. 
It connects several nodes in a peer-to-peer fashion, protecting stored data and communication by using asymmetric encryption. 


\section{Summary and Future work}
\label{sec_summary}

We have sketched the prototype TecCoBot that is able to present automated feedback on writing assignments provided by the software T-MITOCAR.
TecCoBot scales the potential of a special face-to-face mentoring scenario to a large number of students, by enhancing already existing processes, materials and platforms.
This technology-aided support enables providing students a part of mentoring on demand.
Through the use of a software we are able to provide all students individual feedback in time, even in courses with a large number of students and few lecturers. 
Our approach allows students to receive feedback where otherwise they either would have to do without or significant additional human resources would have to be employed.
In this way we make the support for learning from texts in self-study activities available to as many students as possible and whenever they want.
The technologies we have used allow to consider the requirements for privacy, security and informational self-determination.
In the near future we will conduct an application-oriented evaluation regarding the implemented chatbot, as well as reflect about the described processes, which will probably lead to new and adapted research questions. 
Furthermore we will extend the current chatbot prototype with more parts of mentoring.
We also want to collect specific data of students to make technology-aided mentoring more effective and generating knowledge about how it works. Finally we want to identify design concepts to make parts of face-to-face mentoring scalable.


\end{document}